\documentstyle[epsf,12pt]{article}
\newlength{\dinwidth}
\newlength{\dinmargin}
\begin{document}
\input{psfig}
\vspace{1 cm}
\renewcommand{\thefootnote}{\arabic{footnote}}
\def\ycut{y_{\rm cut}}
\def\z{z}
\def\gsubh{\gamma_{H}}
\def\qsd{Q^2}
\def\mjj{\protect\sqrt{\hat{s}}}
\def\mj2{\protect\hat{s}}
\newcommand{\sleq} {\raisebox{-.6ex}{${\textstyle\stackrel{<}{\sim}}$}}
\newcommand{\sgeq} {\raisebox{-.6ex}{${\textstyle\stackrel{>}{\sim}}$}}


\title {
{\bf Jet production in high \mbox{\boldmath $Q^2$} \\
deep-inelastic ep~scattering at HERA}
       \\
\author{\rm ZEUS Collaboration \\}
}
\date{ }
\maketitle
\vspace{5 cm}

\begin{abstract}
Two-jet production in deep-inelastic electron-proton scattering has been
studied for $160<Q^2<1280$~GeV$^2$, $0.01<x<0.1$ and $0.04<y<0.95$ with
the ZEUS detector at HERA.  The kinematic properties of the jets and the
jet production rates are presented. The partonic scaling variables of the
two-jet system and the rate of two-jet production are compared to
perturbative next-to-leading order QCD calculations.
\end{abstract}

\vspace{-20cm}
\begin{flushleft}
\tt DESY 95-016 \\
February 1995 \\
\end{flushleft}

\setcounter{page}{0}
\thispagestyle{empty}

\newpage

%
\def\3{\ss}
\footnotesize
\renewcommand{\thepage}{\Roman{page}}
\begin{center}
\begin{large}
The ZEUS Collaboration
\end{large}
\end{center}
\noindent
M.~Derrick, D.~Krakauer, S.~Magill, D.~Mikunas, B.~Musgrave,
J.~Repond, R.~Stanek, R.L.~Talaga, H.~Zhang \\
{\it Argonne National Laboratory, Argonne, IL, USA}~$^{p}$\\[6pt]
R.~Ayad$^1$, G.~Bari, M.~Basile,
L.~Bellagamba, D.~Boscherini, A.~Bruni, G.~Bruni, P.~Bruni, G.~Cara
Romeo, G.~Castellini$^{2}$, M.~Chiarini,
L.~Cifarelli$^{3}$, F.~Cindolo, A.~Contin,
I.~Gialas, P.~Giusti, \\
G.~Iacobucci, G.~Laurenti, G.~Levi, A.~Margotti,
T.~Massam, R.~Nania, C.~Nemoz, F.~Palmonari, A.~Polini, G.~Sartorelli,
R.~Timellini, Y.~Zamora Garcia$^{1}$,
A.~Zichichi \\
{\it University and INFN Bologna, Bologna, Italy}~$^{f}$
% ...
\\[6pt]
\clearpage
\noindent
\hspace*{1mm}
$^{ 1}$ supported by Worldlab, Lausanne, Switzerland \\
\hspace*{1mm}
$^{ 2}$ also at IROE Florence, Italy  \\
\hspace*{1mm}
$^{ 3}$ now at Univ. of Salerno and INFN Napoli, Italy  \\
\hspace*{1mm}
% ...
$^{12}$ now with OPAL Collaboration, Faculty of Physics at Univ. of
        Freiburg \\
% ...

\begin{tabular}{lp{15cm}}
$^{a}$ &supported by the Natural Sciences and Engineering Research
         Council of Canada (NSERC) \\
$^{b}$ &supported by the FCAR of Quebec, Canada\\
$^{c}$ &supported by the German Federal Ministry for Research and
         Technology (BMFT)\\
$^{d}$ &supported by the MINERVA Gesellschaft f\"ur Forschung GmbH,
         and by the Israel Academy of Science \\
$^{e}$ &supported by the German Israeli Foundation, and
         by the Israel Academy of Science \\
$^{f}$ &supported by the Italian National Institute for Nuclear Physics
         (INFN) \\
$^{g}$ &supported by the Japanese Ministry of Education, Science and
         Culture (the Monbusho)
         and its grants for Scientific Research\\
$^{h}$ &supported by the Korean Ministry of Education and Korea Science
         and Engineering Foundation \\
$^{i}$ &supported by the Netherlands Foundation for Research on Matter
         (FOM)\\
$^{j}$ &supported by the Polish State Committee for Scientific Research
         (grant No. SPB/P3/202/93) and the Foundation for Polish-
         German Collaboration (proj. No. 506/92) \\
$^{k}$ &supported by the Polish State Committee for Scientific
         Research (grant No. PB 861/2/91 and No. 2 2372 9102,
         grant No. PB 2 2376 9102 and No. PB 2 0092 9101) \\
$^{l}$ &partially supported by the German Federal Ministry for
         Research and Technology (BMFT) \\
$^{m}$ &supported by the German Federal Ministry for Research and
         Technology (BMFT), the Volkswagen Foundation, and the Deutsche
         Forschungsgemeinschaft \\
$^{n}$ &supported by the Spanish Ministry of Education and Science
         through funds provided by CICYT \\
$^{o}$ &supported by the Particle Physics and Astronomy Research
        Council \\
$^{p}$ &supported by the US Department of Energy \\
$^{q}$ &supported by the US National Science Foundation
\end{tabular}

\newpage
\pagenumbering{arabic}
\setcounter{page}{1}
\normalsize

\section{Introduction}
\begin{figure}[htbp]
   \centering
   \setlength{\unitlength}{1cm}
   \begin{picture}(15,18)
      \put(0.,0.){\epsfxsize=450pt \epsfbox{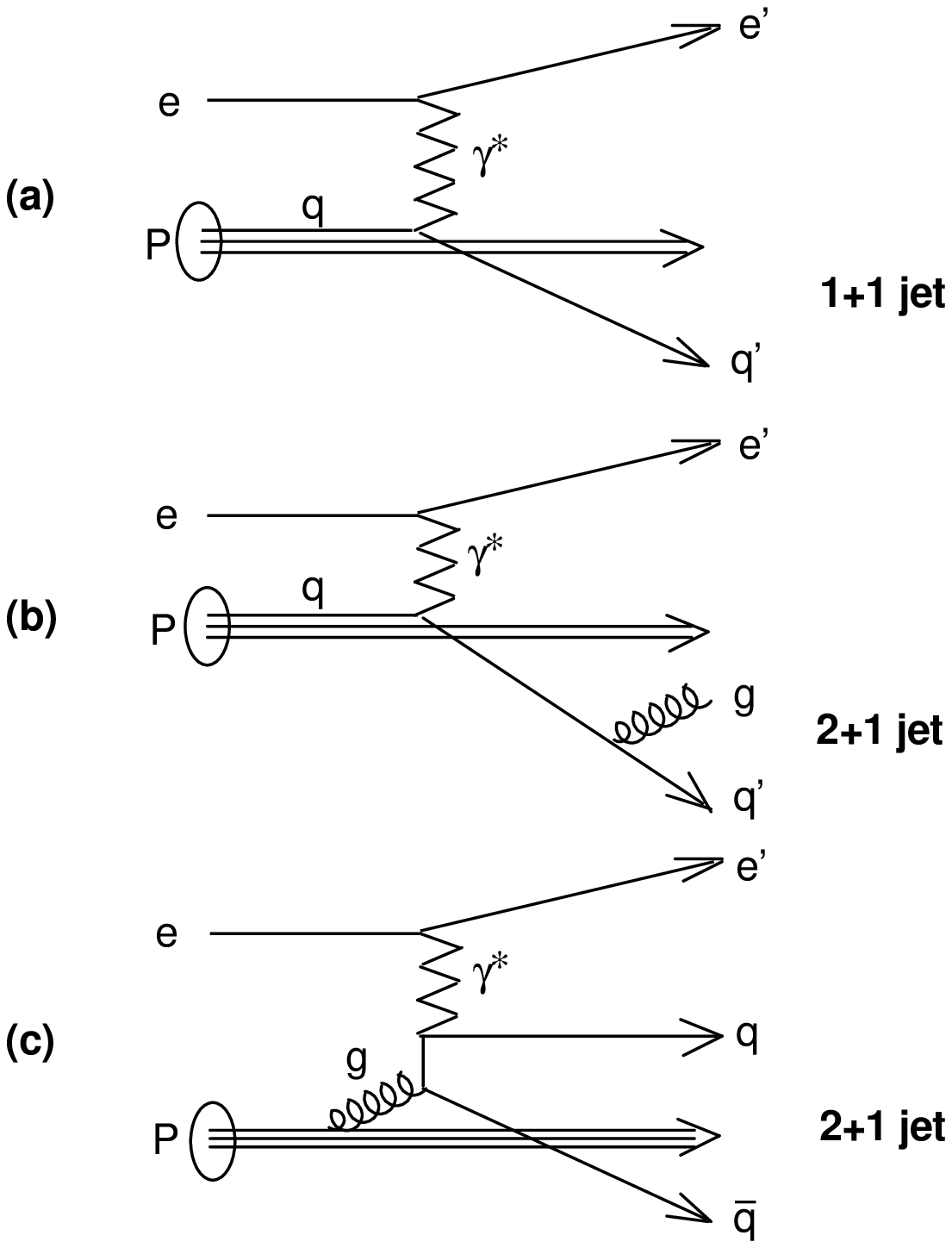} }
   \end{picture}
\caption{Diagrams for neutral current deep-inelastic scattering:
{\bf (a)} Born term,
{\bf (b)}~QCD~Compton scattering,
{\bf (c)} Boson-Gluon-Fusion leading to events with 1+1, 2+1 and 2+1 jets,
vrespectively.
}
\label{feyn}
\end{figure}

Deep-inelastic neutral current scattering is described by
the exchange of a virtual boson ($\gamma^*, Z^0$) between
the electron and a parton in the proton.
In the na\"{\i}ve Quark-Parton-Model
(QPM),
this process leads
to a 1+1 parton configuration in the final state which consists of the
struck quark and the proton remnant,
denoted by ``+1''.
Higher-order
QCD processes contribute
significantly to the ep cross section
at HERA energies: to ${\cal O} (\alpha_{\rm s})$
these are QCD-Compton scattering (QCDC),
where a gluon is radiated by the
scattered quark and Boson-Gluon-Fusion (BGF),
where the virtual boson and a gluon fuse to
form a quark-antiquark pair.
Both processes have 2+1 partons in the final
state, as shown in Fig.~\ref{feyn}.

Jet production in Deep-Inelastic Scattering (DIS) has
been studied by a fixed target experiment (E665)
at a centre of mass energy, $\sqrt{s}$, of 31 GeV and at negative
squared momentum transfers, $Q^2$, of order 10 GeV$^2$ \cite{E665}.
At the  much larger centre of mass energy
of 296~GeV at HERA, jet structures are much more visible
\cite{cone,H1}. It is therefore possible in this energy regime
to determine the $Q^2$ dependence of the
strong coupling constant, $\alpha_{\rm s}$, in a single experiment
by measuring the rate of two-jet\footnote{In this paper we
refer to 2+1 jet production as two-jet production.}
production~\cite{ssr}.

Such measurements require a detailed
understanding of jet properties.
In this paper we study whether
the two-jet system has large enough invariant
mass and whether the jets have sufficiently large transverse momenta for
perturbative calculations to be applicable.
Using Monte Carlo simulations of parton showering and hadronisation to
correct for higher-order and non-perturbative effects,
the underlying parton dynamics and the jet rates are compared to
next-to-leading order (NLO) calculations.

We study jet production in the kinematic range
$160<Q^2<1280$~GeV$^2$, $0.01<x<0.1$ and $0.04<y<0.95$,
at an average squared  hadronic invariant mass $\langle W^2\rangle $ of
14000 GeV$^2$, where $x$ is the ``Bjorken $x$'' variable and
$y$ denotes the energy fraction transferred from the incoming electron
to the proton in its rest system.
The choice of this kinematic range
was based on the following expectations:
at high $Q^2$, jet structures
should be more pronounced and hadronisation
uncertainties should be reduced. Furthermore, in the region $x>0.01$, the
phase space for jet production increases.
Also, at $x>0.01$,  the theoretical uncertainties in the jet rates due
to different parameterisations for the parton densities of the proton
are small \cite{graudenz}.

The data were collected with the ZEUS
detector in 1993 and correspond to an integrated
luminosity of 0.55~pb$^{-1}$.
A brief description of the ZEUS detector is given in
section \ref{zeus}. The event selection is explained
in section \ref{event}. In section \ref{MC}
we describe
the Monte Carlo simulation and the theoretical calculations.
The JADE algorithm~\cite{JADE}
which is used to relate parton- and hadron-level jets in this paper
is described in section~\ref{JADE}.
The kinematic variables for two-jet production are defined
in section \ref{twokin}.
In section \ref{results} we study the hadronic energy
flows, the pseudorapidity distribution of the jets in
the detector, the partonic scaling variables and
the jet production rates.
Conclusions are given in section \ref{conclusion}.

\section{The ZEUS detector}
\label{zeus}
The experiment was performed at the electron-proton collider
HERA using the ZEUS detector. During 1993 HERA operated with
bunches of electrons of energy $E_e=26.7$ GeV colliding with
bunches of protons of energy $E_p=820$ GeV, with a time
between bunch crossings of 96 ns.

ZEUS is a hermetic  multipurpose magnetic detector
that has been described
elsewhere \cite{DIS_ZEUS,MOZART}.
Only components relevant to this analysis are mentioned here.

The hadronic final state and the scattered electron are measured by the
uranium-scintillator calorimeter (CAL).
It consists of three parts, the Forward (FCAL), the
Rear (RCAL) and the Barrel Calorimeter (BCAL)\footnote{
The ZEUS coordinate system is defined as right handed with
the $Z$ axis
pointing in the proton beam direction,
hereafter referred to as ``forward''. The
$X$ axis points towards the centre of HERA, the
$Y$ axis points upward.
The polar angle $\theta$ is taken with respect to the
$Z$ direction.}.
Each part is subdivided longitudinally
into one electromagnetic section (EMC) and one hadronic section (HAC)
for the RCAL or two HAC sections for BCAL and FCAL.
Holes of $20\times 20$ cm$^2$
at the
centre of FCAL and RCAL accommodate the HERA beam pipe.
In the $XY$ plane around the FCAL beam pipe,
the HAC section is segmented in
$20\times 20$ cm$^2$ cells
and the EMC section in $5\times 20$ cm$^2$ cells.
In total, the calorimeter consists of approximately 6000 cells.
In terms of pseudorapidity $\eta=-\ln\tan
\frac{\theta}{2}$, the FCAL covers
$4.3\ge\eta\ge 1.1$, the BCAL
$1.1\ge\eta\ge -0.75$ and the RCAL
$-0.75\ge\eta\ge -3.8$, assuming the nominal interaction point (IP)
at $X=Y=Z=0$.
The CAL energy resolution as measured under test beam conditions
is $\sigma_E/E=0.18/\sqrt{E}$ ($E$ in GeV) for electrons
and $\sigma_E/E=0.35/\sqrt{E}$ for hadrons.
The timing resolution of the calorimeter
is less than 1 ns for energy deposits greater than 4.5 GeV.

The beam monitor scintillation
counter (C5) was used to measure the timing of the
proton and electron bunches.
The event vertices were
determined by drift chambers surrounding the beam pipe:
the Vertex Detector (VXD) and
the Central Tracking Detector (CTD).
The resolution of the $Z$ coordinate of the primary
vertex is 4 mm.

\section{Event selection}
\label{event}
\subsection{General selection}
During the 1993 data-taking period about $10^6$ events
from the DIS trigger branch were
recorded.
A more detailed description of
the DIS trigger conditions and some aspects of the event
selection can be found in \cite{F2}. The trigger
acceptance was essentially independent of the
DIS hadronic final state with an acceptance greater than
97 \% for $Q^2>10$ GeV$^2$.
In order to select DIS
events, the following set of cuts was applied:

\begin{itemize}

\item
      A timing cut required that the event times measured by the
      FCAL and the RCAL were consistent with an interaction inside the
      detector. This cut strongly reduced beam-gas background.

\item
      A scattered electron candidate had to be reconstructed with an
      energy $E_{e'}$
      greater than 10 GeV to ensure high purity of
      the electron sample.

\item
The position $(X,Y)$ of the scattered electron in the RCAL had to lie
outside a square of $32 \times 32$ cm$^2$ centred on the beam axis, to
ensure the electron was fully contained within the detector and its position
could be reconstructed with sufficient accuracy.
\item
Events were selected by the requirement
$35<\delta=
\sum_{i} E_i (1-\cos\theta_i)<60$ GeV,
where $E_i$, $\theta_i$ are the energy and polar angle
(with respect to the nominal IP) of the calorimeter cells $i$.
For fully contained events
$\delta\simeq 2E_e=53.4$ GeV,
where $E_e$ is the electron beam energy.
This cut was applied in order
to remove background due to photoproduction  and beam-gas interactions.
\item   The
$Z$ position of the
event vertex was reconstructed from the tracking data.
Events were accepted
if the $Z$ position was inside
$\pm$75 cm of the nominal IP.

\item Events from beam halo muons, cosmic rays and QED Compton processes
were identified and rejected by suitable algorithms.

\end{itemize}

\subsection{Kinematics}
Because the ZEUS detector is almost hermetic, the kinematic
variables $x$, $y$ and $Q^2$ can be reconstructed in a variety of
ways using combinations of electron and hadronic system energies and
angles \cite{DA}:
\begin{enumerate}
\item The electron method in which
the kinematic
variables are reconstructed from the energy $E_{e'}$ and angle
$\theta_{e'}$ of the scattered electron.
\item The Double Angle ({\it DA}) method in which the angles of the
scattered electron ($\theta_{e'}$) and of the hadronic system
($\gamma_H$) are used. This method reduces the
sensitivity to energy scale uncertainties.
The angle $\gsubh$ corresponds to that of a
massless object balancing the momentum vector of the electron to
satisfy four-momentum conservation.  In the na\"{\i}ve QPM
$\gsubh$ is the scattering angle of the struck quark. It is
determined from the hadronic energy flow measured in the detector
using the equation
\begin{eqnarray}
\cos \gsubh = \frac{(\sum_{h} p_{{\rm X}})^{2}+(\sum_{h} p_{{\rm Y}})^{2}
- (\sum_{h} ( E-p_{{\rm Z}}))^{2}}
{(\sum_{h} p_{{\rm X}})^{2}+(\sum_{h} p_{{\rm Y}})^{2} + (\sum_{h}
( E-p_{{\rm Z}}))^{2}}
\label{cosg}
\end{eqnarray}
where the sums, $\sum_{h}$,
run over all calorimeter cells $h$ which are not
assigned to the scattered electron, and
$(p_{{\rm X}},p_{{\rm Y}}, p_{{\rm Z}})$
is the momentum vector assigned to each cell of
energy $E$.  The cell angles are calculated from the geometric centre
of the cell and the vertex position of the event.
\item The Jacquet-Blondel ({\it JB}) method \cite{JB} in which the kinematic
variables are calculated from the reconstructed hadronic
final state using
the momentum vector $(p_{{\rm X}},p_{{\rm Y}}, p_{{\rm Z}})$
constructed from the energies $E$ and angles
$\theta$ of the calorimeter cells which are not
assigned to the scattered electron.
\end{enumerate}
The electron method
gives better resolution in $x$ at low $Q^2$ while the {\it DA} method is less
sensitive to the calorimeter energy scale and
gives the better mean resolution over the whole $x$--$Q^2$ plane.
The {\it JB} method is used to calculate the visible hadronic energy
in the events which will be used in the jet definition.
Complete
formulae for calculating the variables $x$, $y$ and $Q^2$ are to be found
in \cite{DA}. When it is necessary to distinguish which method
has been used the subscripts `{\it e}', `{\it DA}' or `{\it JB}'
will be used on the variable
concerned.
\newpage
The data sample used for this analysis had to
satisfy the following cuts:
\begin{itemize}
\item
      In the {\it DA} method, in order that the hadronic system be well
      measured, it is necessary to require a minimum of hadronic activity in
      the CAL away from the beampipe.
      For this purpose
      the value of $y_{\rm JB} = \sum_{h}E(1 -\cos\theta) / 2 E_{e}$
      had
      to be greater than 0.04.

\item The value of $y_{e}=1-\frac{E_{e'}}{2E_{e}}
      (1-\cos\theta_{e'})$ had
      to be less than 0.95. This cut rejects photoproduction
      background.
\end{itemize}

The kinematic variables $Q^2$ and $x$ were determined with the
{\it DA} method. Starting from 28200 events with
$Q^2_{\rm DA}>10$ GeV$^2$, 1020 events remain
after selecting the kinematic region \linebreak
$160<Q^2_{\rm DA}<1280$~GeV$^2$ and
$0.01<x_{\rm DA}<0.1$ for this analysis.
The remaining photoproduction background in this high $(x,Q^2)$ data
sample is $\sleq 1$\%.

The corrected
distributions are corrected for detector and acceptance effects and
given in the kinematic ranges
$160<Q^2<1280$~GeV$^2$,
$0.01<x<0.1$ and
$0.04<y<0.95$. To investigate the evolution of
jet structures with $Q^2$, a low
$(x,Q^2)$ data sample of 4230 events is used
with $10<Q^2<20$~GeV$^2$,
$0.0012<x<0.0024$ and
$0.04<y<0.95$.

\section{QCD calculation and event simulation}
\label{MC}
\subsection{LO simulations}
The detector effects were simulated using events generated
with the LEPTO 6.1 Monte Carlo
program \cite{LEPTO} based on the  electroweak
scattering cross section.
We have studied the process $ep\rightarrow e'X$
considering $\gamma^*$ and $Z^0$ exchange.

The leading order (LO) matrix elements (ME) are used to simulate
the QPM process, $\gamma^*+q\rightarrow q$,
the QCDC process, $\gamma^*+q\rightarrow q+g$, and
the BGF process, $\gamma^*+g\rightarrow q+\bar{q}$
(Fig.~\ref{feyn}). In order to
obtain finite cross sections for parton
emissions, a minimum invariant mass,
$m_{ij}^2=y_{\rm min}W^2$,
is required
for all pairs of final state partons $i$ and $j$.
Higher order parton emissions,
which are calculated
in the leading log approximation (LLA)
of perturbative QCD, are
simulated by the Parton Shower (PS) model.
The struck parton can radiate partons
either before
or after the interaction. The amount and the hardness
of the radiation depends on the virtuality (mass)
of the partons.
In order to simulate the hard emission
of partons and the higher-order parton showers,
a combined option (MEPS) exists.
The value $y_{\rm min}=0.015$ was used.
The parameterisation of the parton distribution
functions was the MRSD$'_-$ set \cite{MRSD}
which has been shown to describe reasonably the HERA measurements
of the proton structure function, $F_2$~\cite{F2}.

The event generation included the effects of initial- and
final-state photon radiation which were calculated
with the program HERACLES 4.4 \cite{HERACLES}.
The simulation of the detector used a
program based on GEANT 3.13 \cite{MOZART,GEANT}.

\subsection{NLO calculations}
\label{NLO}
The LEPTO 6.1 Monte Carlo event generator uses
the exact ${\cal O} (\alpha_{\rm s})$ matrix element (ME) and the
parton shower (PS) in the leading log approximation.
It does not include
the NLO matrix element calculation.
However, the NLO corrections to the 2+1 jet cross
section
due to unresolved 3+1 jet events and
due to virtual corrections are
significant \cite{graudenz}.
These corrections are included in
the program DISJET of Brodkorb and Mirkes \cite{DISJET} and
in a similiar
program by Graudenz called PROJET \cite{PROJET}.

Both programs calculate cross sections at the partonic level
in NLO as a function of $x$ and $Q^2$.
PROJET~3.6, in addition, provides cross sections
in terms of the parton variables (see section \ref{twokin}),
but it does not contain the NLO corrections to the
longitudinal cross section\footnote{These
corrections are contained in
a newer version (PROJET 4.1.1).}.
Only the exchange of a virtual photon ($\gamma^*$) is
considered in the calculations, however the contribution
from $Z^0$ exchange is small (expected to be less than 2\%
for $Q^2$ around 1000~GeV$^2$ \cite{F2}).

\section{The jet finding algorithm}
\label{JADE}
A jet finding algorithm is necessary to relate
the hadronic final states measured in the detector to
hard partonic processes.
Results on two-jet production in DIS
have been published previously by the ZEUS collaboration
based on a cone algorithm using a cone of radius
$R=\sqrt{(\Delta \eta)^2+(\Delta\phi)^2}=1$~\cite{cone};
here the cone variables are the pseudorapidity $\eta$ and the
azimuthal angle $\phi$.

In this paper we use
the JADE algorithm \cite{JADE},
since it is currently the only algorithm
which allows comparison to the NLO calculations.
The performance of the JADE algorithm in reconstructing
jets has been compared
to other algorithms
in \cite{hedberg}.
The JADE algorithm
is a cluster
algorithm based on
the scaled invariant mass
$$y_{ij}^{\rm JADE}=
\frac{2 E_i E_j(1-\cos\theta_{ij})}{W^2}$$
for any two objects $i$ and $j$
assuming that these objects are massless.
$W^2$ is the squared invariant mass
of the hadronic final state and $\theta_{ij}$ is the angle
between the two objects of energies $E_i$ and $E_j$.
The minimum $y_{ij}$
of all possible combinations is found. If
the value  of this minimum $y_{ij}$  is less than
the cut-off parameter $\ycut$, the two objects $i$ and
$j$ are
merged into a new object by adding their four-momenta and the process is
repeated until all $y_{ij}>\ycut$. The surviving
objects are called jets.

The JADE algorithm is applied at the parton, hadron and
detector levels in the HERA laboratory frame.
At the detector level, the calorimeter cell energies and
positions serve as inputs to the JADE algorithm.
For this analysis,
the scale parameter used in the JADE algorithm, $W^2$, is taken
at the detector level
from just those calorimeter cells
associated with the hadronic system,
\mbox{$W_{\rm JB}^2=s(1-x_{\rm JB})y_{\rm JB}$.}
By using
$W_{\rm JB}$, the detector acceptance corrections
for $y_{ij}$ largely cancel.

We have found that
the JADE scheme as it is described here leads to smaller
hadronisation corrections than the
Lorentz invariant $E$-scheme \cite{LEP} which uses
the exact invariant mass
$m_{ij}$ for massive objects. For this analysis, the JADE scheme
is used in the HERA laboratory frame.

The losses in the forward beam pipe of the ZEUS detector are
taken into account by
adding a fictitious cluster  (called
pseudo-particle) in the forward direction
to which the missing longitudinal
momentum for each event is assigned. The pseudo-particle is treated
like any other particle in the JADE clustering
scheme.
No pseudo-particle procedure is
used for the parton and
hadron levels of the Monte Carlo generator.

\section{Two-jet kinematics}
\label{twokin}
The differential cross section for two-jet production
depends on five independent
kinematic variables \cite{zerwas}, here taken
as $x,Q^2,x_p,\z$ and $\phi^*$.
The variable $\phi^*$ is the azimuth between the outgoing parton
plane and the lepton scattering plane
in the $\gamma^*$-parton centre of mass
system (CMS) and the variables $x_p$, $\z$ are
Lorentz invariant partonic scaling variables.

The event variable $x_p$ is determined by:
$$x_p=\frac{Q^2}{2p\cdot q}=\frac{x}{\xi}
\,\,\,\,\,\,\, (0\le x_p\le 1),$$
where $\xi$
is the fraction of the proton's (longitudinal) momentum $P$ carried
by the incoming parton of momentum $p$ ($p=\xi P$) and $q$ is the
momentum of the exchanged virtual boson ($Q^2\equiv -q^2$).
Alternatively, one can write:
$$x_p =\frac{Q^2}{Q^2+\mj2}$$
for massless jets, where
$\mjj$
is the invariant mass of the two-jet system.
This expression is used to determine $x_p$ experimentally.
The range of values of $x_p$
is given by $x$
as the lower limit
and is a function of $\ycut$ at the upper limit:
$$x =
 \frac{Q^2}{Q^2+ W^2}
\le x_p \le
 \frac{Q^2}{Q^2+\ycut W^2}. $$
The jet scaling variable $\z$ is
related to the angular distribution of the jets
in the $\gamma^*$-parton CMS:
$$\z = \frac{P\cdot p_{\rm jet}}{P\cdot q}=
\frac{1}{2}\left(1-\cos\theta^*_{\rm jet}\right)
\,\,\,\,\,\,\, (0\le \z \le 1),$$
where $\theta^*_{\rm jet}$
is the polar angle of the jet
in the $\gamma^*$-parton CMS.
$\z$ is measured  from the jet, which is assumed to be massless,
and reconstructed in the HERA system:
$$\z
=\frac{E_{\rm jet}(1-\cos\theta_{\rm jet})}{\sum_{i}
E_i(1-\cos\theta_i)}
$$
where the sum runs over the two reconstructed (massless) jets.
In the JADE algorithm, the total hadronic energy is contained in
the 2+1 jet system.
Since the remnant jet (`+1') has no transverse momentum,
$z_{\rm remnant}=0$ and therefore $z_{1}+z_{2}=1$.
The minimum value is determined by the value of $\ycut$ and is given by:
$$\z_{min} =\frac{(1-x)x_p}{x_p-x}\ycut\equiv 1-\z_{max}.$$

In this analysis, we integrate over $\phi^*$ and
study the dependence of the two-jet production
on $x_p$ and $\z$.
The transverse momentum $P_T$
of the jets with respect to the $\gamma^*$ direction
in the $\gamma^*$-parton CMS, can be
derived from $x_p$, $\z$ and $Q^2$:
$$P_T=\sqrt{Q^2 \frac{\z}{x_p}(1-x_p)(1-\z)}.$$
At the ${\cal O}(\alpha_{\rm s})$
tree level, the singularities in the two-jet cross section
are given by \cite{korner}:
$$d\sigma_{2+1}^{\mbox{\protect\footnotesize BGF}}\propto
\frac{[\z^2+(1-\z^2)][x_p^2+(1-x_p^2)]}{\z(1-\z)}\;\; \mbox{and}\;\;
d\sigma_{2+1}^{\mbox{\protect\footnotesize QCDC}}
\propto
\frac{1+x_p^2z_{q}^2}{(1-x_p)(1-z_{q})},$$
where $z_q$ is labelled specifically for the quark jet in the QCDC process.
The $x_p$ singularity in the QCDC process results in small jet-jet invariant
masses being preferred.
Both processes have a singularity at $\z= 0$ or 1:
in the BGF process, it
is related to the collinear emission
of the two quarks and, in the QCDC process, to the collinear
or soft emission of the gluon.
The kinematic properties of the JADE jets and the relationship
to these ${\cal O}(\alpha_{\rm s})$
singularities are studied in section \ref{xpzp}.

\section{Results}
\label{results}
\subsection{Transverse energy flow in jets}
\label{flow}
The jet properties were studied at a
fixed $\ycut$ of 0.02. This
value is chosen as a compromise between the increase of the
higher-order corrections at lower $\ycut$ values
and the loss of
statistics at higher $\ycut$ values for two-jet events.
This choice resulted in 237 events with 2+1 jets
from a total number of 1020 events in the high $(x,Q^2)$ region.
The distributions in the following sections are normalised
to the total number of events, $N_{ev}$.

\begin{figure}[htbp]
   \centering
   \setlength{\unitlength}{1cm}
   \begin{picture}(15,18)
      \put(0.,0.){\epsfxsize=450pt \epsfbox{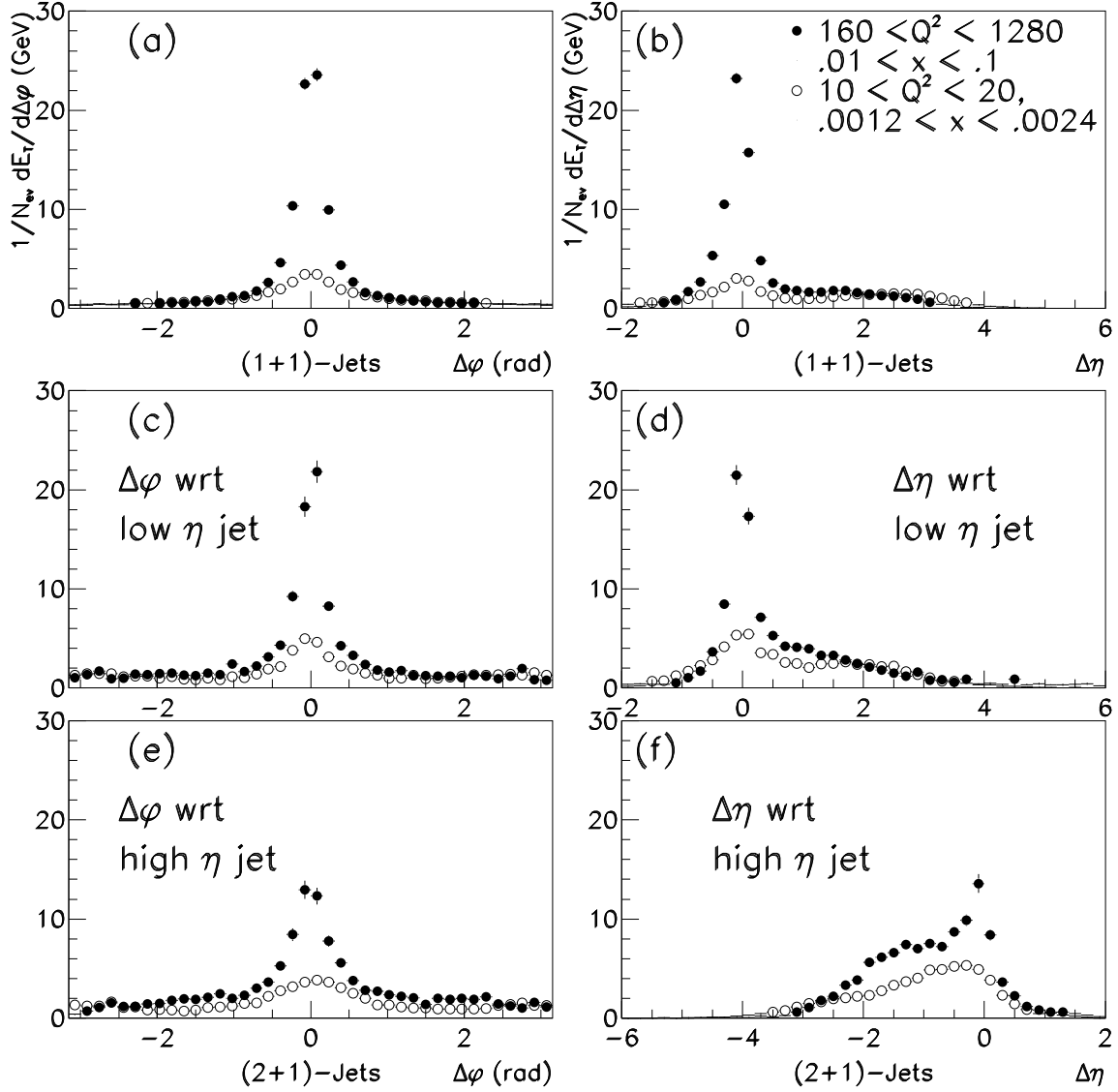} }
   \end{picture}
\caption{Transverse energy flows in the HERA frame measured
relative to the jet directions
in 1+1 {\bf (a--b)} and 2+1 {\bf (c--f)} jet events for
two ($x,Q^2$) intervals
($10<Q^2<20$ GeV$^2$, $0.0012<x<0.0024$) and
($160<Q^2<1280$ GeV$^2$, $0.01<x<0.1$).
The data are uncorrected and are indicated by the full and open dots,
respectively.}
\label{et_top}
\end{figure}

\begin{figure}[htbp]
   \centering
   \setlength{\unitlength}{1cm}
   \begin{picture}(15,18)
      \put(0.,0.){\epsfxsize=450pt \epsfbox{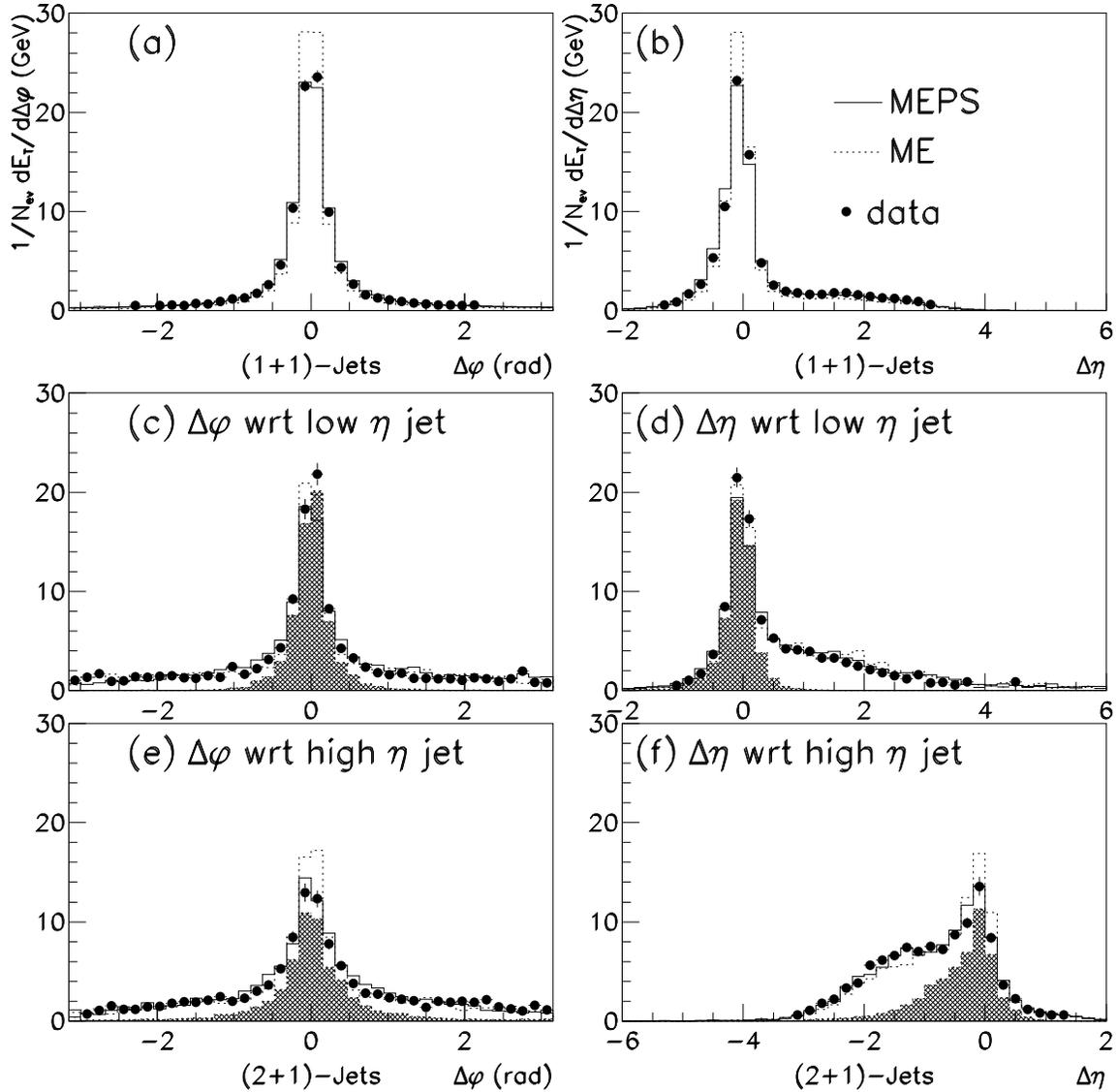} }
   \end{picture}
\caption{Transverse energy flows in the HERA frame measured
relative to the jet directions
in 1+1 {\bf (a--b)} and 2+1 {\bf (c--f)} jet events for
$160<Q^2<1280$ GeV$^2$ and $0.01<x<0.1$.
The data are shown by the full dots.
The contribution from the jet which defines
$\Delta\eta=0$ or $\Delta\phi=0$ for the data
is indicated by the shaded histogram.
Uncorrected data are compared to the MEPS and the ME simulations.}
\label{et_jet_fin}

\end{figure}

The transverse energy ($E_T$) flows relative to
the $Z$ axis in the laboratory frame
$$ \frac{1}{N_{ev}}\frac{dE_T}{d(\Delta\eta_i)} \;\;\
\;\mbox{and}\;\;\;
 \frac{1}{N_{ev}}\frac{dE_T}{d(\Delta\phi_i)}$$
\noindent were calculated with
respect to the jet directions in the
laboratory frame by defining
$$\Delta\eta_i=\eta_i-\eta_{\rm jet} \;\; \mbox{and}
\;\;\Delta\phi_i=\phi_i-\phi_{\rm jet}$$
for every calorimeter cell $i$ of 1+1 jet events
(Figs.~\ref{et_top}a,b and \ref{et_jet_fin}a,b)
and separately relative to the axes of the higher and the
lower $\eta$ jets of 2+1 jet events
(Figs.~\ref{et_top}c--f and \ref{et_jet_fin}c--f).

In Fig.~\ref{et_top}, the $E_T$ flows are compared in two
different kinematic regions at low $Q^2$
($10<Q^2<20$ GeV$^2$, $0.0012<x<0.0024$) and at high $Q^2$
($160<Q^2<1280$ GeV$^2$, $0.01<x<0.1$).
The average $E_T$ flow increases with $Q^2$ and
the jets are more collimated at higher $Q^2$.
{}From kinematics
$\langle E_T\rangle\simeq Q$ for the electron and
therefore the same is expected for the balancing 1+1 jet. For the
two kinematic regions shown here, this corresponds
to approximately a factor 5 increase in $\langle E_T\rangle$.
The influence of the proton remnant
is visible as a tail at $\Delta\eta>1$ in the 1+1 jet events
(Fig.~\ref{et_top}b), whereas
the tail in Fig.~\ref{et_top}d is mainly
assigned to the second jet and not the proton remnant.

Fig.~\ref{et_jet_fin} shows
the $E_T$ flows for the high $(x,Q^2)$ region
compared to two Monte Carlo data
sets after the detector simulation,
generated with the ME and the MEPS options.
The $E_T$ flow in terms of the energy scale and
the shape around the jet direction is well described,
especially when using the MEPS option.
In the ME model, the absence of parton showering
results in the $E_T$ flow being more concentrated around the jet direction.

Every cell contributing to the
$E_T$ flow is assigned to one of the jets by the JADE algorithm.
The $E_T$ flow contribution
of the jet defining
$\Delta\eta=0$ or $\Delta\phi=0$
is
shown as the shaded histogram  in Fig.~\ref{et_jet_fin}.
The JADE algorithm is observed to typically assign cells to a jet
inside the range $|\Delta\eta|\sleq 1$, however, the algorithm also
assigns cells beyond this range for the higher $\eta$ jet
(Fig.~\ref{et_jet_fin}e-f).

\subsection{Properties of two-jet events}
\label{sec-xp}
\begin{figure}[htbp]
   \centering
   \setlength{\unitlength}{1cm}
   \begin{picture}(15,18)
      \put(0.,0.){\epsfxsize=450pt \epsfbox{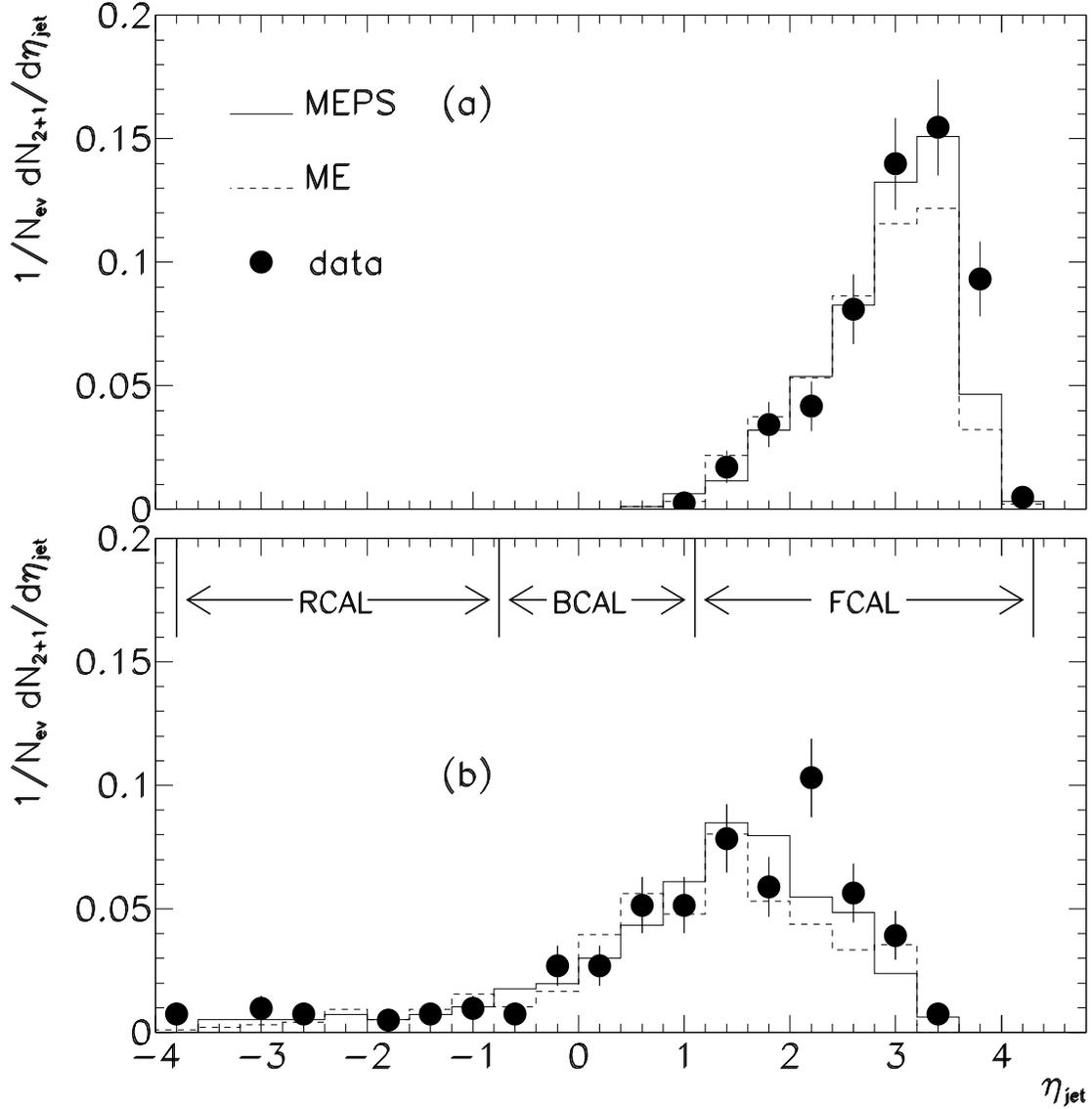} }
   \end{picture}
\caption{Pseudorapidity $\eta_{\rm jet}$
of the two jets in the kinematic ranges
$160<Q^2<1280$~GeV$^2$, $0.01<x<0.1$ and $0.04<y<0.95$:
{\bf (a)} higher $\eta$ jet;
{\bf (b)} lower $\eta$ jet.
Uncorrected  data are compared to the MEPS and the ME simulations.
The boundaries of the different calorimeter parts are indicated.}
\label{eta}
\end{figure}

The two-jet properties are studied in this section
for the high $(x,Q^2)$ region.
In Fig.~\ref{eta},
the distribution of the pseudorapidity
$\eta_{\rm jet}$  of the two jets is
shown. The jets are ordered in $\eta$.
The higher $\eta$ jet is usually found
very close to the forward beam pipe.
About half of the jet axes are reconstructed in cells within 30~cm
of the FCAL region near the beam pipe ($\theta\sleq 8^\circ$).
In this forward region, the results depend on
the description of the initial state parton shower and
of the target fragmentation in the
Monte Carlo generator, as well as on the simulation of
the response of the calorimeter around the beam pipe.

Fig.~\ref{eta}a
shows that the predictions of
the $\eta_{\rm jet}$ distribution by the ME and the MEPS
models describe the data fairly well
except for the very forward angles ($\eta_{\rm jet}>3.6$)
where the predictions are below the data
for both models.
In the region $3.6>\eta_{\rm jet}>2.8$
the data are better described by the MEPS model.
Most of the lower $\eta$ jets are also found at positive
pseudorapidities in the BCAL or in the FCAL.
The Monte Carlo model predictions for
the $\eta_{\rm jet}$ distribution of the lower $\eta$ jet, which
is usually well separated from the beam pipe region,
gives in general a good description of the data (Fig.~\ref{eta}b).

\begin{figure}[htbp]
   \centering
   \setlength{\unitlength}{1cm}
   \begin{picture}(15,18)
      \put(0.,0.){\epsfxsize=450pt \epsfbox{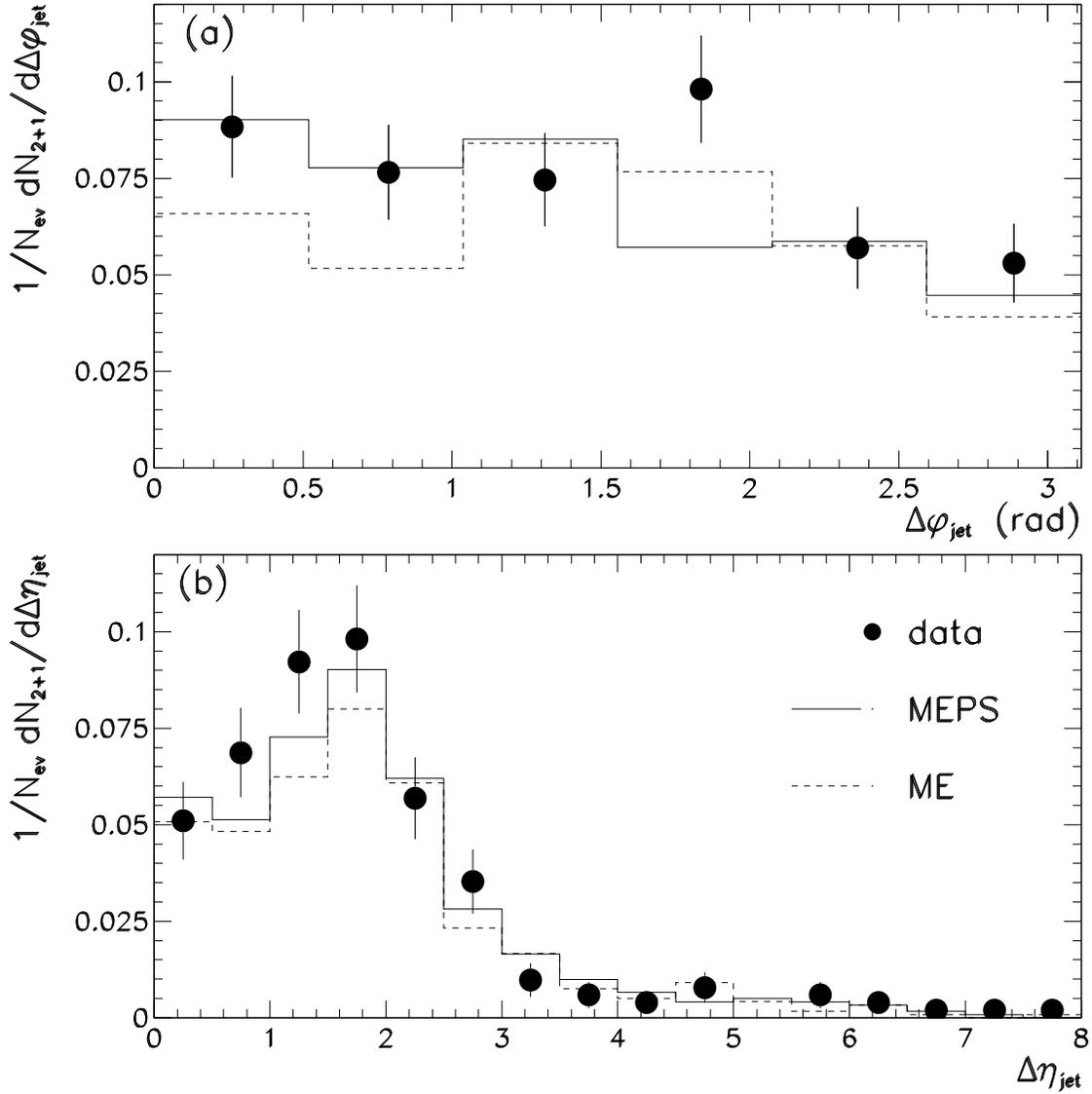} }
   \end{picture}
\caption{{\bf (a)} The difference in
azimuthal angle, $\Delta\phi_{\rm jet}$, and {\bf (b)}
pseudorapidity, $\Delta\eta_{\rm jet}$,
between the two jets of 2+1 jet events in the kinematic ranges
$160<Q^2<1280$~GeV$^2$, $0.01<x<0.1$ and $0.04<y<0.95$.
Uncorrected  data are compared to the MEPS and the ME simulations.}
\label{dr}
\end{figure}

The distributions of the differences in
azimuthal angle, $\Delta\phi_{\rm jet}$,
(Fig.~\ref{dr}a) and
pseudorapidity, $\Delta\eta_{\rm jet}$,
(Fig.~\ref{dr}b)
between the two jets
show that the jets found by the clustering scheme of
the JADE algorithm are reasonably well separated
in $\eta$. Since $\phi$ is
calculated in the laboratory frame and the electron
acquires more transverse momentum at larger values
of $Q^2$, the two jets are not back-to-back in
the laboratory frame for the high $(x,Q^2)$ data.
This differs from low-$Q^2$ jet production, where
the jets are typically back-to-back in $\phi$.
The data are generally well described by the Monte Carlo
predictions for
$\Delta\eta_{\rm jet}$ and $\Delta\phi_{\rm jet}$.

\begin{figure}[htbp]
   \centering
   \setlength{\unitlength}{1cm}
   \begin{picture}(15,18)
      \put(0.,0.){\epsfxsize=450pt \epsfbox{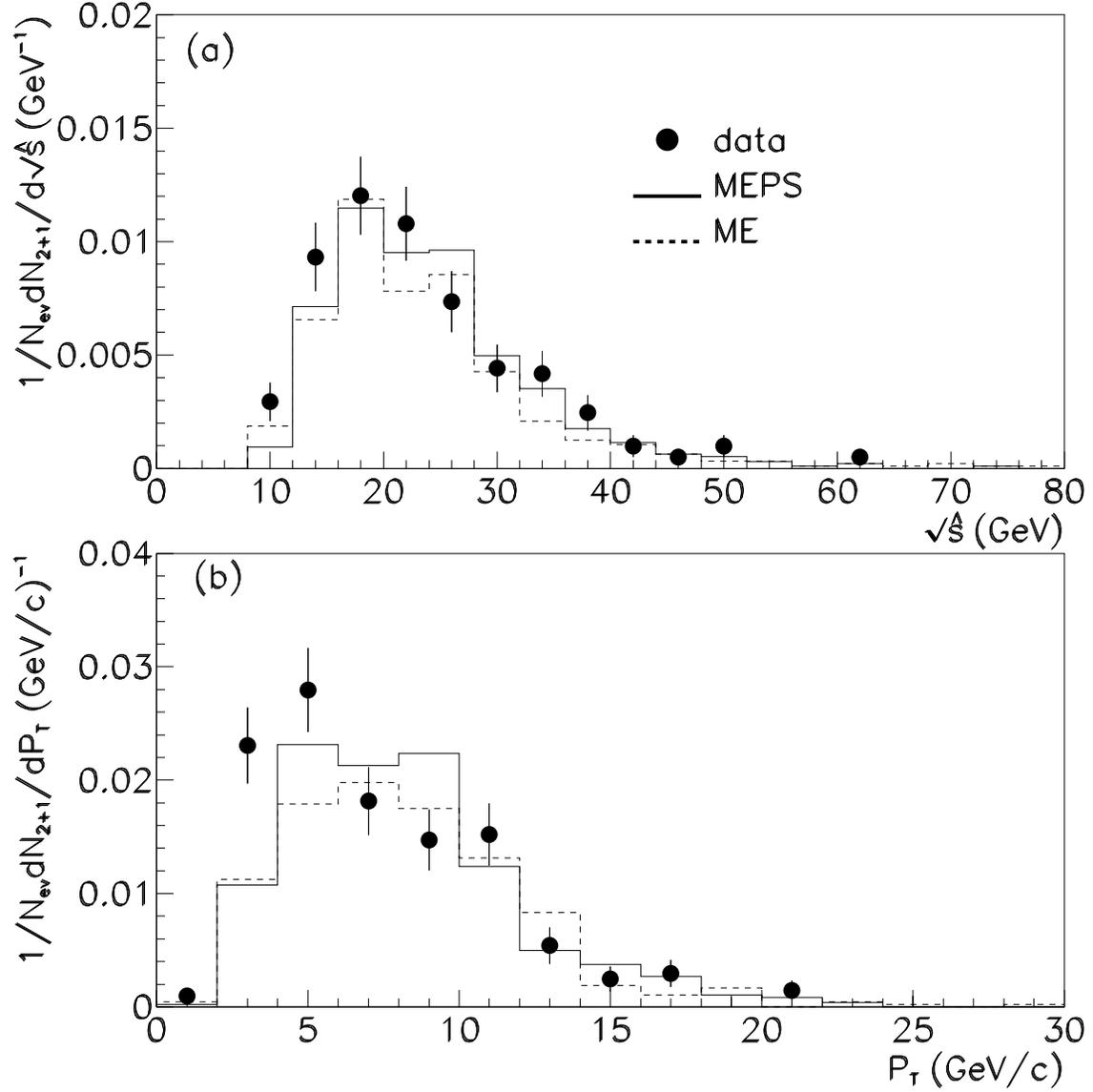} }
   \end{picture}
\caption{{\bf (a)} The two-jet invariant mass $\mjj$ and {\bf (b)}
the jet transverse momentum $P_T$
in the $\gamma^*$-parton centre of mass system
in the kinematic ranges $160<Q^2<1280$~GeV$^2$,
$0.01<x<0.1$ and $0.04<y<0.95$.
The uncorrected data are compared to the MEPS and the ME simulations.}
\label{m2}
\end{figure}

The invariant mass $\mjj$ of the two jets is shown in Fig.~\ref{m2}a.
The average invariant mass
$\langle \mjj \rangle$ of the jets is about 23 GeV.
The $\mjj$ distribution is reasonably well described
by the two  models.
The $P_T$ of the jets in the $\gamma^*$-parton CMS,
determined from the expression in section~6, is
shown in Fig.~\ref{m2}b.
(Only the $P_T$ distribution of one of the two jets is
shown, since $|P_T^{(1)}|$=$|P_T^{(2)}|$.)
The average value $\langle P_T \rangle $ is about 7~GeV/c.
This is sufficiently large to ensure the validity of a
perturbative QCD calculation.
At low $P_T$, however, the data lie above the Monte Carlo prediction.

\subsection{Partonic scaling variables}
\label{xpzp}
\begin{figure}[htbp]
   \centering
   \setlength{\unitlength}{1cm}
   \begin{picture}(15,18)
      \put(0.,0.){\epsfxsize=450pt \epsfbox{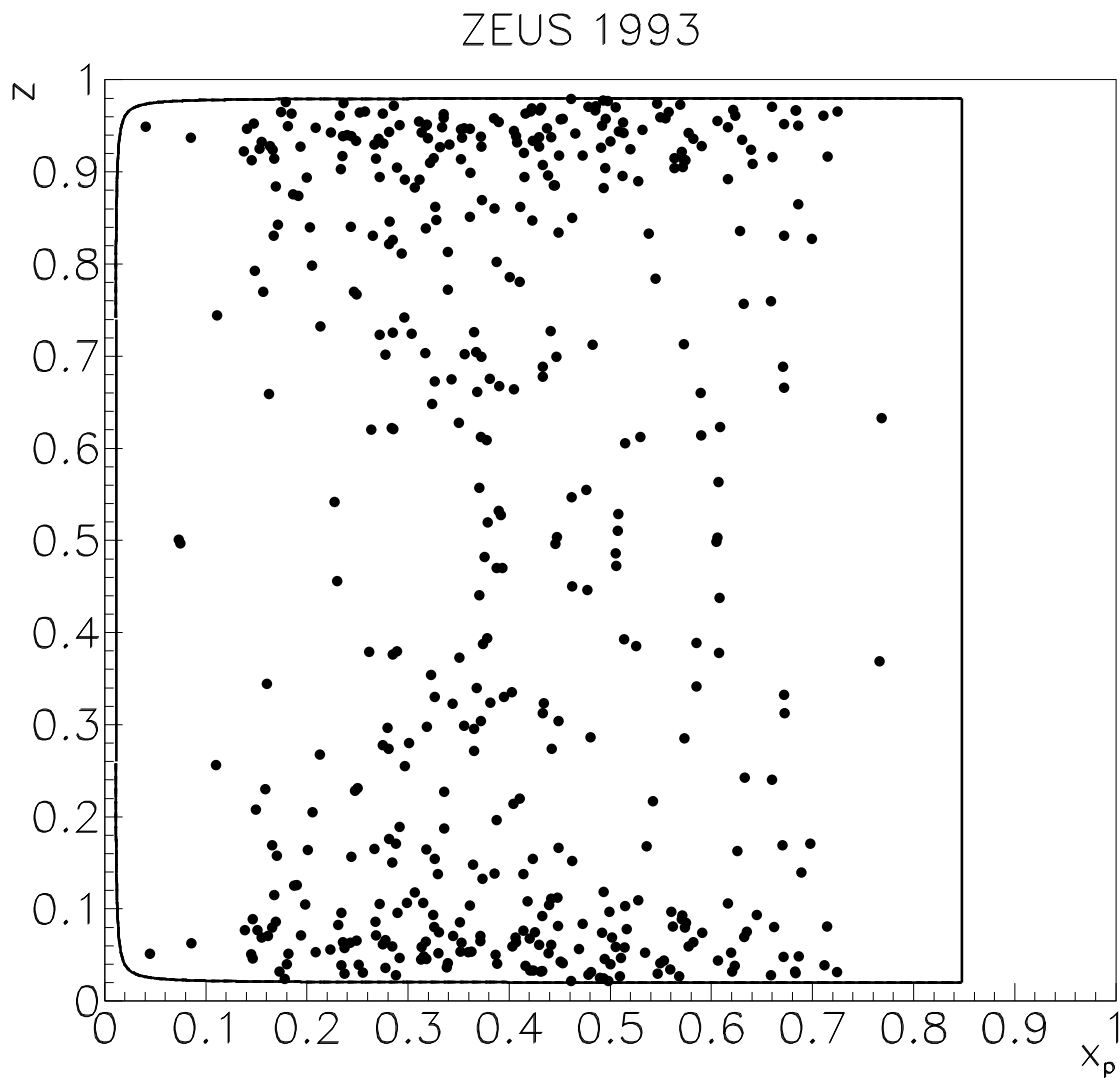} }
   \end{picture}
\caption{The distribution of $\z$ versus
$x_p$ for uncorrected data in the high $(x,Q^2)$ interval.
The area defined by the curve indicates the
kinematic limit for $\ycut=0.02$ in the range $0.01<x<0.1$.
The limit on $\z$ is given by $z_{min}\simeq\ycut=1-z_{max}$.
}
\label{walsh}
\end{figure}

The distribution of $\z$ versus
$x_p$ for the uncorrected data
is presented in
Fig.~\ref{walsh}.
The area defined by the curve
specifies the
kinematic limit for $\ycut=0.02$  in the JADE algorithm
and $0.01<x<0.1$.
The upper and lower limits
on $x_p$ depend on $x$. The upper limit varies
in the range $0.34<x_p<0.84$
for $0.01<x<0.1$, respectively.
The data typically lie close to the $\z$ limit, given
by $z_{min}\simeq\ycut=1-z_{max}$, which
is close to the singular regions of the cross section
discussed in section~6.

\begin{figure}[htbp]
   \centering
   \setlength{\unitlength}{1cm}
   \begin{picture}(15,18)
      \put(0.,0.){\epsfxsize=450pt \epsfbox{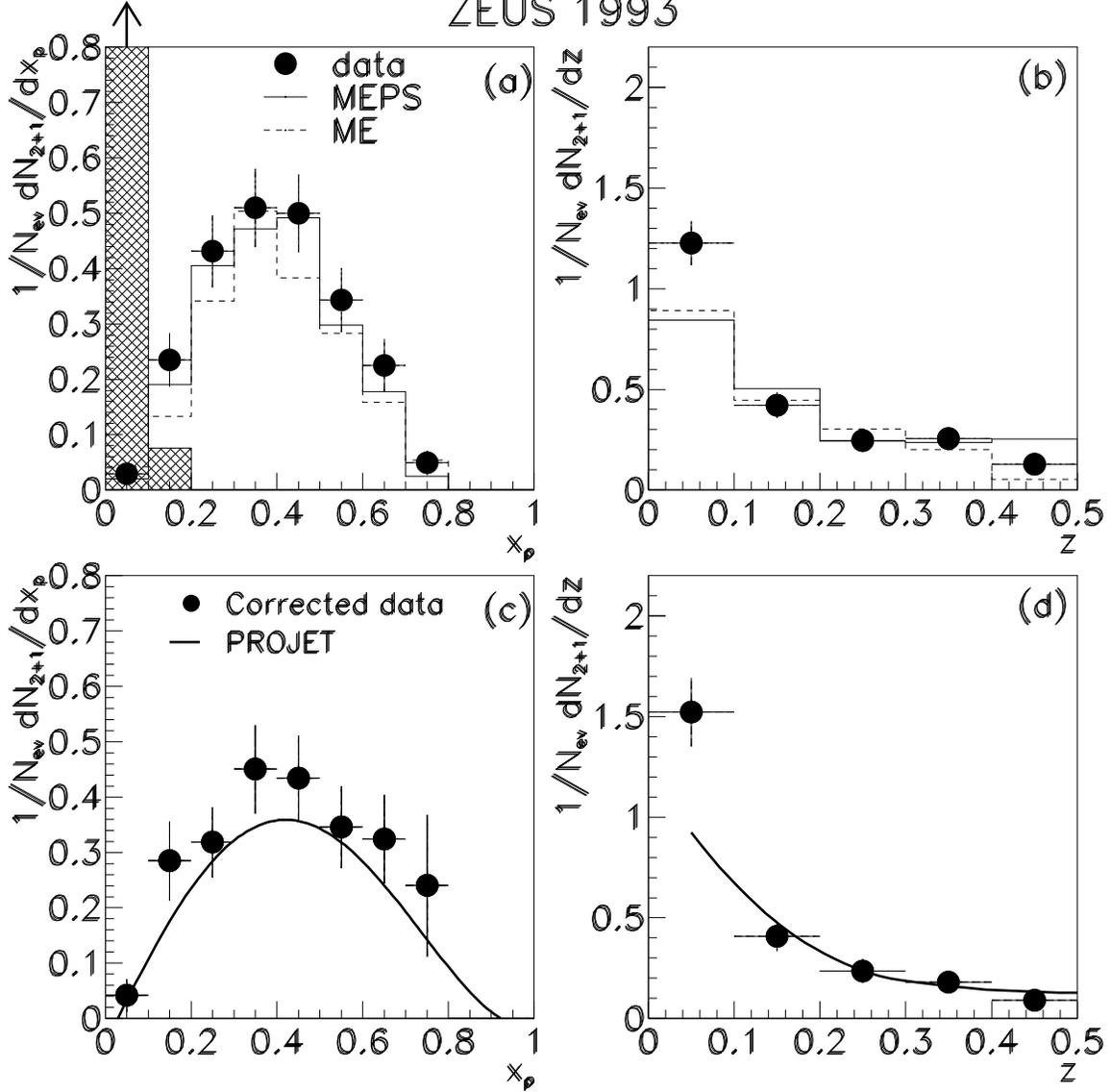} }
   \end{picture}
\caption{{\bf (a)} Uncorrected $x_p$ distribution of two-jet events and
{\bf (b)} uncorrected $\z$ distribution of the jet with the smaller $\z$.
The uncorrected data are compared to
the MEPS and the ME simulations.
The shaded histogram in {\bf (a)} shows the uncorrected $x_p$ distribution in
the kinematic ranges
$10<Q^2<20$ GeV$^2$, $0.0012<x<0.0024$ and
$0.04<y<0.95$.
\newline
{\bf (c)} corrected $x_p$ distribution and
{\bf (d)} corrected $\z$ distribution
compared to the NLO calculation (PROJET).
The data have been corrected with the MEPS simulations.
Statistical errors only are shown.
}
\label{xp_zp}
\end{figure}

The measured $x_p$ distribution for the two-jet events is shown
in Fig.~\ref{xp_zp}a at the detector level.
The bin size has been chosen to reflect the
experimental resolution of around 10\% in $x_p$.
In the considered $(x,Q^2)$ region, there exists almost no phase
space limitation at the upper end of the $x_p$ distribution,
with $x_p$ extending up to approximately 0.8 for
the high $Q^2$ data.
This is however sufficiently far from the $x_p$ singularity
of the QCDC process discussed in section~\ref{twokin}.
The average $x_p$ is about 0.5,
i.~e.~$\langle \mj2 \rangle \simeq\langle Q^2\rangle$.
The $x_p$ distribution is well described by the MEPS and the
ME models.

The uncorrected $x_p$ distribution for the low $Q^2$ data
which is shown for comparison shows a very different
behaviour. It is peaked at small values of $x_p$,
i.~e.~$\langle \mj2 \rangle > \langle Q^2\rangle$,
due to the $\ycut$ requirement.
At these lower $Q^2$ values, the largest scale
is therefore $\mj2$.

In the uncorrected $\z$ distribution in Fig.~\ref{xp_zp}b,
only the $\z$
distribution for the jet with smaller $\z$ is shown, since
$z_{1}+z_{2}=1$.
At the detector level, the distribution is reasonably well described
by both the ME and the MEPS model, but a discrepancy exists
for $\z<0.1$ which corresponds to jets close
to the forward beam pipe.

The corrected $x_p$ and $\z$ distributions
are compared to the PROJET NLO calculation in
Figs.~\ref{xp_zp}c and d.
These distributions were corrected for detector and
hadronisation effects with the MEPS simulation.
The MRSD$'_-$ parton density
parameterisations \cite{MRSD} are used in the calculations
of PROJET.
Apart from the parameterisation of the parton densities,
the QCD calculations contain only one
free parameter, the strong coupling constant
$\alpha_{\rm s}$.
For the calculations in Fig.~\ref{xp_zp}
a value of $\Lambda^{(5)}_{\overline{\rm MS}}=250$ MeV
was chosen which corresponds to $\alpha_{\rm s}(M_Z^2)=0.120$,
as measured from hadronic event shapes, energy correlations and
jet rates in e$^+$e$^-$ annihilations \cite{LEP}.

The shape of the corrected $x_p$ distribution and the PROJET
calculation is in good agreement, however, the rate is underestimated.
The corrected $\z$ distribution is well-described by the calculation for
$\z>0.1$, however, an excess is observed in data relative to the PROJET
calculation in the lowest $\z$ bin.
This is, however, the region with the largest systematic uncertainties.
The significance of this deviation cannot currently be estimated due to the
experimental and theoretical problems associated with the jets in the forward
region. The $\z$ distributions are almost identical
for the ME(LO), not shown, and the PROJET(NLO) calculation.
The cross section rises as $\z\rightarrow 0$,
because of the low cut-off with respect to the $\z$ singularity
in the JADE algorithm.

It should be noted that the jets at the MEPS parton level
acquire mass in the JADE algorithm through
multiple combinations of partons from the parton shower (PS)
whereas the PROJET and the ME jets
are defined by  massless partons.
This effect shifts both the $x_p$ and the $\z$ distribution
as defined in MEPS,
if $x_p$ and $\z$
are not calculated in the approximation of massless parton jets.

It should also be noted that the kinematic cut-offs in terms
of the partonic scaling variables strongly depend
on the choice of jet algorithm.
In comparison to the JADE algorithm
the $K_{\perp}$ algorithm \cite{kt}
leads to a much less
peaked $\z$ distribution.
The $K_{\perp}$ algorithm was not used for this analysis,
because the NLO calculations for DIS
are based only on the JADE scheme.

\subsection{Jet rates}
\label{rates}
The two-jet rate $R_{2+1}$ is defined by the ratio
$$ R_{2+1}=\frac{N_{2+1}}{N_{2+1}+N_{1+1}}, $$
where $N_{2+1}$ and $N_{1+1}$ are the number of
2+1 or 1+1 jet events
and $R_{1+1}~\equiv~1-R_{2+1}$.
This definition differs from the ``usual''
definition where the denominator is the total
cross section including 3+1 and higher-order
contributions. It is used to reduce the dependence on
the 3+1 jet rate, which is only calculated at
the tree level in PROJET and DISJET. Experimentally,
the acceptance correction factors for the 3+1 jet rate
are large. For $\ycut= 0.02$ about 4\% of all events
have 3+1 reconstructed jets. Almost no
events with 0+1 jets are found
in this high $(x,Q^2)$ interval
whereas they are often found in the lower $(x,Q^2)$ intervals.

\begin{figure}[htbp]
   \centering
   \setlength{\unitlength}{1cm}
   \begin{picture}(15,18)
      \put(0.,0.){\epsfxsize=450pt\epsfbox{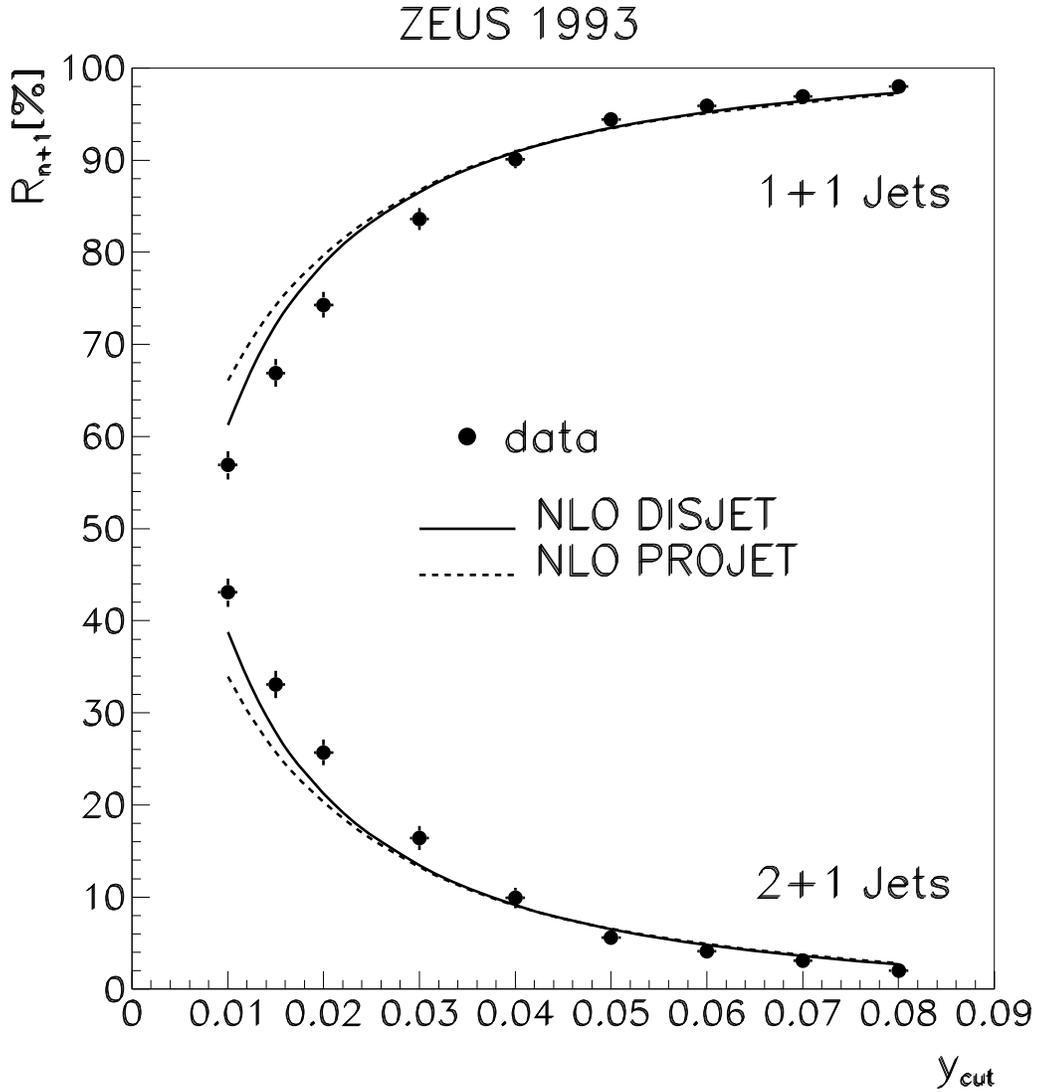} }
   \end{picture}
\caption{The corrected jet production rate $R_{n+1}$
in \% as a function of $\ycut$ is compared to the NLO
calculation of the programs PROJET \protect\cite{PROJET} and DISJET
\protect\cite{DISJET}
in the kinematic ranges
$160<Q^2<1280$~GeV$^2$,
$0.01<x<0.1$ and
$0.04<y<0.95$.
The data have been corrected
to the partonic level with the MEPS model.
Statistical errors only are shown.
}
\label{rj}
\end{figure}

In Fig. \ref{rj},
the corrected jet production rates are shown as a function of
the jet resolution parameter $\ycut$.
The correction  to the parton level
is done using the MEPS model.
The 2+1 jet rate increases with finer
jet resolution (smaller $\ycut$).
The measured jet rates have been corrected
using a bin-by-bin correction method,
where the correction factors are up to 20\% depending on $\ycut$.
At lower $\ycut$ values ($\ycut\simeq 0.02$)
the resolution on $y_{ij}$ is approximately~0.01.

For comparison a full NLO calculation performed with
the DISJET program and the NLO calculation
of the PROJET program are also shown.
The difference between the two calculations is
attributed to the missing NLO corrections to the longitudinal
cross section in PROJET~3.6.
The PROJET and DISJET curves are calculated using
$\Lambda^{(5)}_{\overline{\rm MS}}=250$ MeV and the MRSD$'_-$
parton density parameterisations, as discussed in section~\ref{xpzp}.
Choosing other currently available parameterisations
\cite{pdf} leads to a variation of the
theoretical curves which is of the same
order as the statistical errors on the data.
The errors shown
are the purely statistical binomial errors which are highly correlated,
because all 2+1 jet events at a given $\ycut$ are
included in the points at smaller $\ycut$.
The data agree with the theoretical calculations at the 15\% level, however,
a deviation from the QCD models
at $\ycut$ values below 0.04 is evident in Fig.~\ref{rj}.
This deviation is correlated with the excess for $\z<0.1$ discussed in
section~\ref{xpzp}.

\section{Conclusions}
\label{conclusion}
The production of 2+1 jet events as defined by the JADE
algorithm has been studied in deep-inelastic neutral current
events at HERA for $160<Q^2<1280$~GeV$^2$,
$0.01<x<0.1$ and $0.04<y<0.95$.
In this kinematic range, prominent jet structures
have been observed.
The transverse energy flows with respect to the jet
direction are well described by the LEPTO 6.1 Monte Carlo
program. Various measurements of the kinematic properties
of the jets have been analysed. They are
generally well described by the Monte Carlo models.
The invariant mass and the transverse momentum
of the two jets are large enough to allow
a description in terms of perturbative QCD.
For the first time in DIS,
the partonic scaling variables $x_p$ and $\z$
have been reconstructed from the jets
and are shown to be well described by
NLO calculations for $z>0.1$. Jet rates, corrected to
the parton level, have been measured as a function of
$\ycut$ and compared to NLO calculations.
In addition to the structure function parameterisation,
the QCD calculations have one free parameter,
$\alpha_{\rm s}$, which was taken from
LEP measurements. The sensitivity on the choice of parameterisation is
small in this kinematic regime.
The dynamics of jet production in DIS are satisfactorily described
at the 15\% level by the calculations.
We have refrained at this stage from extracting $\alpha_{\rm s}$
from our data because of the observed sensitivity at very small
values of $\z$ and the as yet incomplete understanding of this region.

\section*{Acknowledgements}
We thank the DESY directorate for their strong support and
encouragement, and the HERA machine group for their
remarkable achievement in providing colliding ep beams.
We also gratefully acknowledge the support of the DESY computing and network
services.

We would like to thank T.~Brodkorb, D.~Graudenz
and E.~Mirkes for valuable discussions and for providing
the NLO calculations.


\end{document}